\documentclass{article}

\def\be{\begin{equation}}
\def\bea{\begin{eqnarray}}
\def\ee{\end{equation}}
\def\eea{\end{eqnarray}}

\title{\bf The thunder of distant Net storms}
\author{Marcelo O. Magnasco \\
Center for Studies in Physics and Biology\\
The Rockefeller University\\
1230 York Avenue, New York NY10021, U.S.A.
}
\begin{document}
\maketitle

\begin{abstract}
\baselineskip 14pt
Computers and routers on the Internet send each other
error messages (called {\bf ICMP datagrams}) to signal conditions such
as network congestion or blackouts. While these datagrams are {\em
very} rare, less than 0.001\% of total traffic, they hold very
important {\em global} information about problems and congestions
elsewhere in the Net. A measurement of the flow of such
error messages in our local cluster shows a very pathological
distribution of inter-message times: $P(\Delta t) \approx 1/\Delta t$.
This scaling extends for about seven decades, and is only punctuated by
extraneously periodic signals from automatons. More than a half of 
these error messages were themselves generated erroneously. 

\end{abstract}

The decentralized and anarchich structure of the Internet has, for a
long time, invited people to perform all sorts of measurements.
Indeed, it has taxed and even challenged our ability to make such
measurements: for instance, it's been several years since we last knew,
to any accuracy, how large the Internet actually is
\cite{HOWLARGE,NWSURV}.  Many types of measurements have shown
``anomalous'' or ``pathologic'' statistics, meaning that the
distribution of some quantity has power-law (a.k.a. Pareto or
``heavy'') tails . A power-law tail implies that the probability of an
outlier diminishes very slowly with the size of the outlier, and hence
measurements show strong irregularities.  While anomalous from a
classical statistics viewpoint, this is a very usual phenomenon in the
natural sciences, and has been extensively studied
\cite{MANDEL2,ONEOVERF}; in particular, one of the earliest examples
concerns noise in communication channels \cite{MANDEL}. As applied to
the Internet, local traffic measurements at the datagram level
\cite{LELWIL,FOWLEL,LTWW,LTWWE,ABRVEI}, session data \cite{PAXFLO},
webserver workloads \cite{WEBDIS,ARLWIL}, USENET thread length
\cite{GUTO} and Web surfing patterns \cite{HUBER2} all show these
tails. There is ample evidence that these statistical pathologies do
not come from intrinsic instabilities of the Net as a communications
system, but rather stem {\em from the way in which people use the
Net}.  Power-law distributions are often regarded as prima-facie
evidence of some self-organizational process, but they can also simply
be a reflection of other underlying power laws. The sizes of all files
on a computer follow a power-law distribution, from small files a few
characters in size, to large datasets (e.g., videos) hundreds of
megabytes in size. It is thus natural that, for instance, the sizes of
documents retrieved from webservers reflect this underlying breadth
\cite{ARLWIL}, and it's also been shown that individual surfers follow
anomalous Levy-flight-like surfing patterns \cite{HUBER2}. A useful and
clarifying distinction was made in \cite{PAXFLO}: the initiations of
``sessions'' (such as telnet or ftp) follow perfectly normal Poissonian
patterns, while the individual transactions within a session (the
ftpdata connections, or the TCP traffic {\em within} a given telnet
session) fluctuate over many scales.

However, all such measurements of local ``traffic'' have built-in
limitations as to how much they can fluctuate. Indeed examination
of the data shows a fairly complicated picture: the bulk of the traffic
is not ``self-similar'' or pathological; only some tails show power-law
scaling over a limited range. See Figure 1. Even a moderate amount of
scaling behaviour can certainly create a lot of trouble for engineers
(who have to design equipment to handle such contingencies); but it is
not pervasive enough, nor is it defined over a large enough range, to
qualify as ``self-similar'' behaviour in the sense in which it is
usually used in the natural sciences. We could say that we are trying
to make a measurement of precipitation during a storm: it will probably
fluctuate and be gutsy, but in its bulk be statistically regular.

We will do something different here: we will not focus on what we can
measure on a storm locally, but rather try to listen for echos of
storms elsewhere. We will do so by focusing on a negligible portion of
the local traffic: the ICMP error messages. These are packets that
computers or routers send to each other on the internet to signal all
sorts of traffic problems: {\em ``speak slower, you're breaking up''}
or {\em ``you can't get there from here''} or {\em ``this bridge is
backed up''}, etc. ICMP error messages are generated only when packets
are ``dropped'', i.e., typically during congestions or blackouts, or
simply to signal that a certain ``place'' does not exist and hence you
can't get there. They are, thus, an indication of a deluge {\em
elsewhere}, the locally received sound of faraway storms \cite{STORMS}.

We measure ICMP error datagrams exchanged between our cluster and the
outside of Rockefeller University. Error messages are rather rare, and
hence it takes a very long time to accumulate a reasonable amount of
events:  our measurement accumulated 11118 datagrams in a span of 29
days.  A histogram of packet inter-arrival times ($\Delta t_i = t_{i+1}
- t_i$) shows a distribution $P(\Delta t) \approx 1/\Delta t$; the
scaling holds for about 7 decades, from 0.3 milliseconds to an hour:
see Figure 2. It is only punctuated by what appear as ``Dirac
deltas'':  bursts with high periodicity embedded within the stream of
error messages.  These bursts typically correspond to ``repeat
transmissions'' of the same error message. The distribution is much
broader than traffic measurements, as shown in Table 1.

\begin{table}
\hrule
\caption{Distribution Widths. \label{table1}}
\vskip 2mm
{\centerline{\begin{tabular}{|ll|rr|r|c|}
\hline
Where& What& 25\% & 75\% & Width & Figure\\
\hline
R.U.    & Full Traffic      & 1.170 & 3.154 & 2.70 & 1a\\
BellCore& Full Traffic      & 1.108 & 3.760 & 3.39 & 1b\\
R.U.    & External Traffic  & 44.891 & 261.276 & 5.82 & 1c\\
BellCore& External Traffic  & 2.524 & 22.372 & 8.86 & 1d\\
R.U.    & ICMP Errors, ext  & 239.511 & 32000    & 133.6 & 2a\\
\hline
\end{tabular}}}
\vskip 2mm
We define the {\em width} of a distribution as the quotient between the
75\% percentile and the 25\% percentile: it thus indicates the range
within which the middle half of the probability is contained. The
value for a Poisson point process is $\log(1/4)/\log(3/4)=4.82$, 
and the value for a Pareto distribution ($G(x)=1-({a\over x})^b$)
is $3^{1/b}$. The width is not translation invariant (as the standard
deviation), but it is scale invariant; unlike the standard deviation,
it is always defined. Times for the percentiles in milliseconds.
\vskip 1mm\hrule
\end{table}

The periodics peaks observed also show, on detailed inspection, some
intriguing features. The peaks are listed in Table 2. Several peaks are
exceedingly thin; for instance, the 64 seconds peak has a half-width of
1 millisecond. However, most of the events in this peak were generated
by a router immediately outside the ``walls'' of Rockefeller, just
three hops away, which explains the timing accuracy.  Not all peaks are
that easily explained. The 24 seconds peak also has a half-width of 1
millisecond; the 84 events within the half-width were generated by 48
different routers and leaf nodes; most of them are in excess of 18 hops
away, with latencies about 500ms, and some of them are as far away as
Germany, France and England; 19 of them were not on the DNS name
tables.

\begin{table}
\hrule
\caption{Peaks \label{table2}}
\vskip 2mm
{\centerline{
\begin{tabular}{|rr|rcrr|cl|}
\hline
$\Delta t$ (sec) & $\pm$ h.w. &\#&$\to$&R&L&DU&Comments \\
\hline
0.240&0.001 &87&out&3&5& port & RealAudio timeouts\\
0.500&0.010 &79&in&40&8& host & Nearby Router \\
1.000&0.010 &51&in&34&8&  & \\
1.500&0.010 &103&out&86&8& port & \\
2.000&0.002 &68&in&13&5& port &  \\
(5.5&  $\to$ 6.0)\qquad & 977&in&336&16&All & Heterogeneous\\
24.000&0.001&84&in&50&4& & far routers\\
32.000&0.002&12&out&1&1&port& frgn DNS timeout\\
64.000&0.001&27&in&1&3&host& nearby router.\\
75.000&0.003&22&in&14&2&host& far routers\\
\hline
\end{tabular} }}
\vskip 1mm
\hrule
\end{table}

It is interesting to note that $P(t)\approx 1/t$ is a highly anomalous
scaling, a limiting case of Pareto distributions. Not only does it lack
standard deviation and mean, it is not even normalizable in the absence
of {\em both} short time and long time cutoffs. In fact, it is the only
power law which shows a symmetry under exchange of short and long
times: both are {\em equally} divergent. Using the transformation $\nu
= 1/t$ and $P(t) dt = P(\nu) d\nu$, we get $P(\nu) = 1/\nu$ because
$d\nu = dt/t^2$, so our measurement shows a discrete point process
version of $1/f$ noise. Thus, the distribution of times between
consecutive error messages displays a behaviour which has {\em no free
parameters at all}, except for the cutoffs. No features or details of
the engineering underneath are left in this background, no traces of
any of the many protocols.

Recently, Huberman and Lukose \cite{HUBER} analysed the issue of {\em
global} jamming on the internet. They were able to show that a simple
game of cooperation and defection by Internet users would cause global
jams, even when they used a trivial network transport model and a very
simplified model of individual activity. The jams, though, have a
nonpathological waiting time distribution, perhaps because the
transport model used ignored the highly structured, tree-like
connectivity of the Internet. This connectivity structure can be shown
to provide a bias towards $1/t$ behaviour in the kind of measurement
reported here.  As seen from a given computer (a ``leaf node''), the
jumble of routers connecting to the rest of the Internet looks pretty
much like a tree, though there are a few cycles due to the return paths
from routers and multiple routes to a given destination. This tree
changes with time, as the structure of routing changes; but at any
given time, it still looks pretty much like a tree. Counting how many
routers are exactly $n$ hops away from us, we observe the number to
roughly double each hop \cite{HOPCOUNT}, so modeling the net as a
binary tree, while a gross oversimplification, is nevertheless somewhat
metrically correct; the {\em real} tree has hugely varying
connectivity. If a link goes down for whatever reasons, it will take a
while until nearby routers learn how to route around it. An attempt to
access a host routed through this link will result in a ``Destination
unreachable'' error message.  Assuming a nonpathological distribution
(links go down and up as uncorrelated processes with identical
probabilities per link), attempts to randomly access the leaves
generates a $1/t$ distribution of waiting times between error messages,
see simulation output on Figure 3.

This model is, however, not enough to explain our data. Though the 
basic power law is the same, the cutoffs strongly disagree. The extent
of the scaling region is determined by the number of levels in the
tree; in order to fit this extent to our data, the Internet should
have several billion nodes. The lowest cutoff is given by the frequency
at which the root node attempts to access leaves; in order to match our
lower cutoff of $<1$ ms, our domain would have to be attempting to access
thousands of different computers per second, which is orders of magnitude
wrong. The model should thus not be understood as an explanation of $1/t$
behaviour in our data, but rather as a basic structural bias towards
power-law behaviour that the geometry imposes on the system. 

In a classic study, S. Bellovin \cite{PACKETS} described a ``natural
computer virus'': a DNS cache-corruption virus. Hosts on the
Internet rely upon DNS name servers to get the address of a computer
based on its name, and viceversa; DNS servers communicate with one
another to get this information. In addition, DNS servers will "cache"
the names they've obtained from other nameservers for a while. If this
cache or a portion thereof is corrupted, then the DNS server will serve
incorrect addresses, not just to hosts, but to other DNS servers, thus
propagating the errors. It is highly likely that DNS is not the only
Internet service that can support such self-propagating entities; there
might indeed be a veritable ecosystem evolving on the backwaters of the
net. Interestingly, evolutionary pressure on such potential beings
would be to stay out of sight, since their being noticed would result
in fixing of the software bug that allows them to propagate in the
first place. The generic feature enabling such a state of confusion to 
propagate is the fact that a master server cannot determine that it is
confused, and can't reply "I am or might be confused, ask someone else". 
Thus the server keeps speweing confusion. Quite a few services have
such structure.

\begin{table}\hrule
\caption{Breakout by Protocol\label{table3}}
{\centerline{
\begin{tabular}{|r|rl|rl|}
\hline
Packets& &Type& &Code \cite{RFC1700}\\
\hline
   1 & 3& Dest unreachable & 2& Bad protocol\\
   4 &11& Time exceeded & 1& in reassembly\\
  18 & 3& Dest unreachable & 10&Forbidden host\\
  21 & 3& Dest unreachable & 4& Needed to fragment\\
  21 & 3& Dest unreachable & 9&Forbidden net\\
  23 & 3& Dest unreachable & 0& Bad net\\
  86 & 5& Redirect & 0& network \\
 256 & 5& Redirect & 1& host \\
 983 & 4& Source Quench & &{\bf Incorrect}$^1$ \\
1665 &11& Time exceeded & 0& in transit\\
2206 & 3& Dest unreachable & 3& Bad port\\
2497 & 3& Dest unreachable & 1& Bad host\\
3300 & 3& Dest unreachable & 13&Communication\\
& & & & administratively forbidden\\
\hline
\end{tabular}
}}
\vskip 2mm
$^1$ RFC1812 \cite{RFC1812} stipulates that {\em Source Quench} messages should
not be sent.
\hrule
\end{table}

As applied to our case, we have been measuring {\em errors}, 
and most of them are generated by {\em routers}, and routers pass
to one another information that is deemed to be authoritative through
RIP. So it would be interesting to
assess whether the ICMP errors are correctly generated or are, themselves,
erroneous. Table 3 shows a breakout by protocol of our data set. A fair
fraction of the errors seen there should not have been sent at all;
{\sl source quench} datagrams, for instance, have been deprecated for a
long time. The {\sl redirect} messages should not have made it out of
their respective local net, and the {\sl redirect host} class consists
exclusively of messages like ``to reach host {\bf x}, use {\bf x} as a
gateway'', which are obviously nonsensical \cite{PACKETS}. {\sl Time
exceeded in transit} are typically generating through erroneous routing
loops \cite{PAX96}.  But the most frequent class of errors, {\sl
Destination unreachable}, poses an interesting problem: since these
problems are typically transient, how can one evaluate whether the
datagrams were sent correctly or in error? A way out is to notice that
the {\em most} frequent subclass of errors is {\sl communication
administratively forbidden}. This should mean the host is in a
protected subnet, typically behind a firewall, a condition which is
definitely not transient. We generated a list \cite{TRAILER} of all the
hosts to which an access attempt had resulted in this error in our
dataset, and attempted access all over again. We got through without
errors to 152 hosts out of 320 ($\approx 50\%$), which means that these
hosts were not placed behind administrative restrictions. These hosts
were responsible for 2297 error datagrams out of the 3300 ($\approx
70\%$).

Thus we can say that, in all likelihood, {\bf more than a half
of our data set consists of error messages which were themselves
generated erroneously}, through incorrect routing. Obviously, router
confusion dominates over actual physical errors. These states of
confusion could last substantially longer than actual link or host
downtimes, and perhaps even spread or self-organize in some fashion.
This may be an underlying cause of power-law organization in the data. 
Let us also recall that a fundamental law of error correction circuitry
design is that the error correction components are just as fallible as
the rest of the circuit. Parity correction algorithms have to assume the
error might very well be in the parity bit, for instance. This fundamental
notion has not been implemented in communication protocols like RIP or
ICMP: there is no ICMP message reading "I'm confused".

As use of the Internet continues to spread, and as research and
educational institutions prepare for the Next Generation Internet and
Internet2 projects, it becomes more important to understand the 
global, large scale dynamics of our world-wide network. We've shown
that important clues to this dynamics may lie in rather insignificant
fractions of the overall traffic of the Net. 

I would like to thank G. Cecchi, D. Chialvo, J.-P. Eckmann, M.
Feigenbaum, A.  Libchaber and E. Siggia. This work was supported in
part by the Sloan Foundation.


\subsection*{Figure Captions}
\qquad

{\bf Figure 1:} Cumulative probability distributions $G(\Delta t)$
of packet interarrival times for four different traffic measurements:
(a) 150000 packets of local
traffic at our cluster, (b) one million packets of local traffic at Bellcore
(c) 150000 packets of external traffic at out cluster, (d) one million 
packets of external traffic at Bellcore. Traces (a) and (c) were 
measured locally for this study; traces (b) and (d) were studied
in \cite{LELWIL,FOWLEL,LTWW,LTWWE} and are publicly available \cite{TRF}. 

{\bf Figure 2:} (a) Cumulative probability distribution $G(\Delta t)$ and
probability density function $P(\Delta t)$ for the interarrival
times of ICMP error datagrams between our cluster and the outside
of the Rockefeller campus. Notice that the probability density 
follows $P(\Delta t) \approx 1/t$ for about 7 decades, except for 
sharp Dirac-$\delta$-like peaks. 
(b) Probability density for a two month trace over two subnets.

{\bf Figure 3:} Distribution of inter-event times in a simulation of
network access errors due to down links on a binary tree. The tree
had 29 levels, so there were $2^{29}$ leaf nodes and $2^{29}-2$ routers
and links. At any given time 100 links are down; they stay down for 
an average of $10^5$ iterations. 


\begin{thebibliography}{99}
\bibitem{HOWLARGE} Recent surveys \cite{NWSURV} show about 30 million 
net addresses have been ``publicly'' assigned, and more than 5 million
computers respond to ``public'' queries such as ping (ICMP echo request). 
However, many domains hide their computers (and sometimes their names)
from the public (behind firewalls, for instance), and many ``home'' 
machines connected through ISPs are only actually connected a fraction of the 
time, so the actual number of computers that connect to the net 
is probably quite larger. We performed an estimate of addresses 
replying to ping by pinging random addresses (all four numbers randomly
generated between 0 and 255): 80 out of 38791 attempts replied, 
giving us an estimate around 9 million machines replying to ping. 
35 did not have names in DNS, making this estimate consistent with 
\cite{NWSURV}. This estimate has strong variations with the time of day, 
indicating substantial amounts of machines which go on- and off-line
during the day.
\bibitem{STORMS} Very low frequency electromagnetic radiation (``VLF
atmospherics'') carries EM radiation from lightning discharges all over
the globe. This ``radio thunder'' has $1/f$ burstyness patterns, so our
analogy is not entirely random.
\bibitem{TRAILER} The ICMP error datagram includes both the address of the 
host generating the error (in the header) and the host to which the datagram
spawning the error was addressed (in the trailer); we made a list of the 
latter ones.
\end{thebibliography}
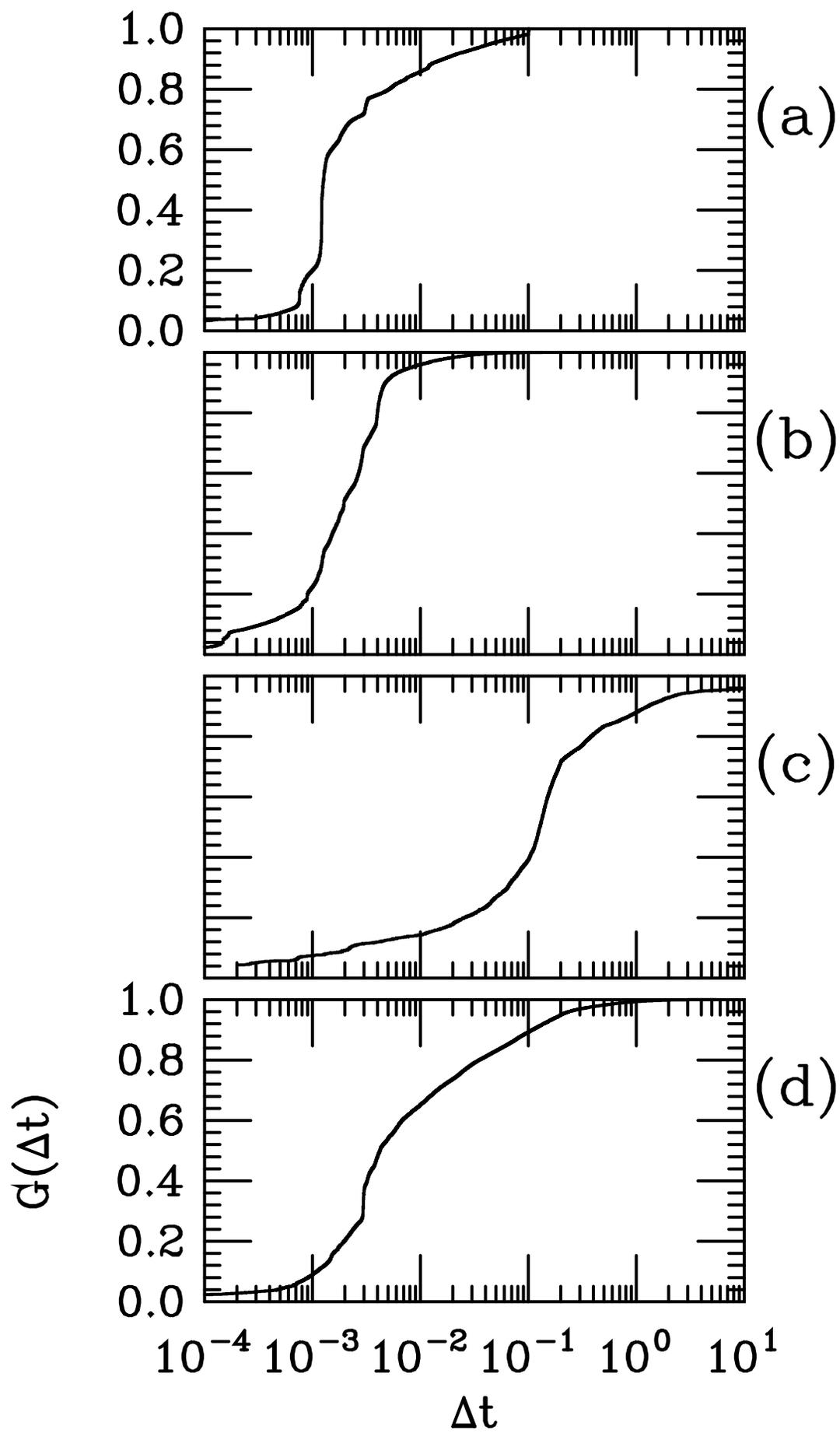
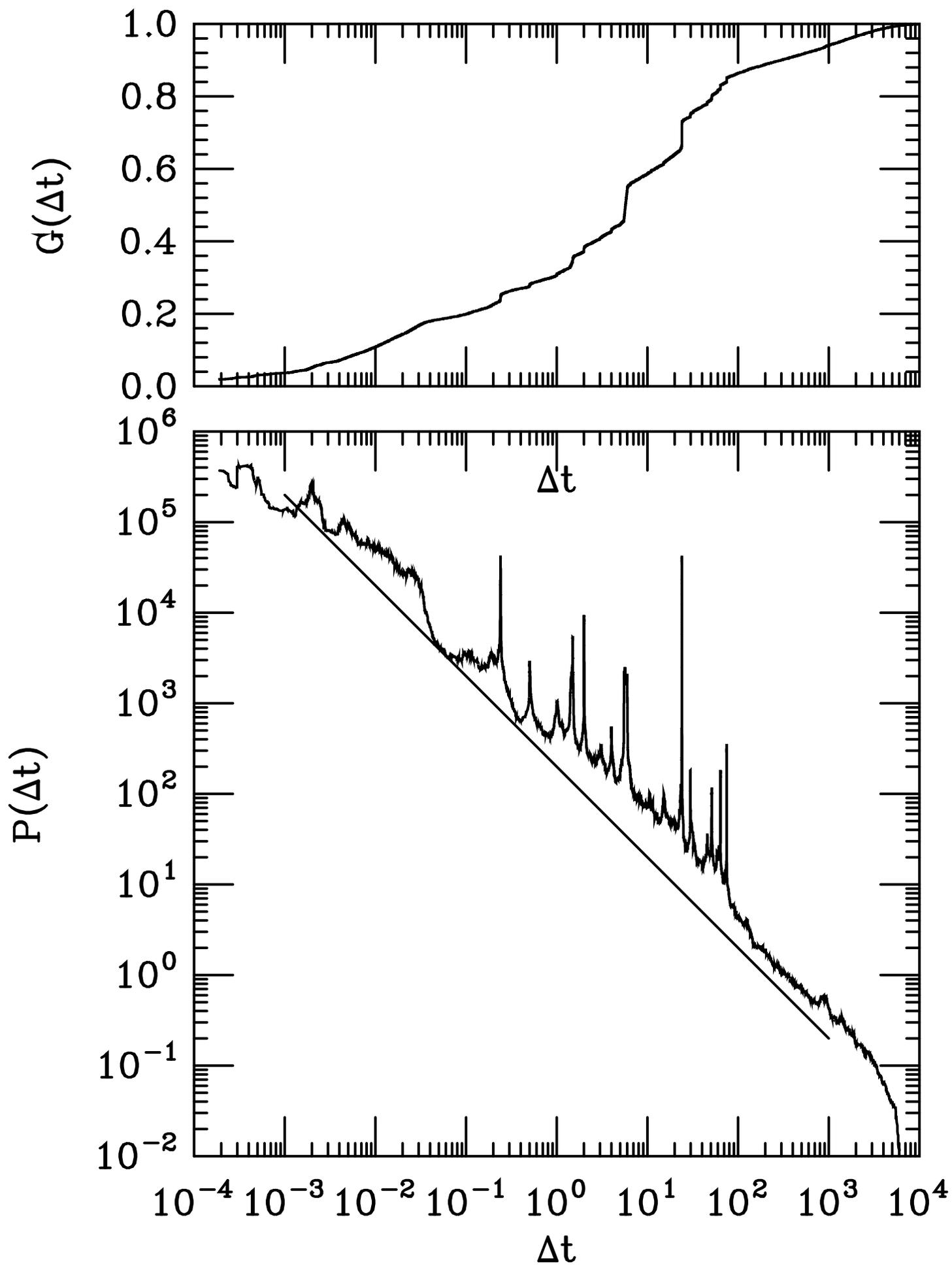
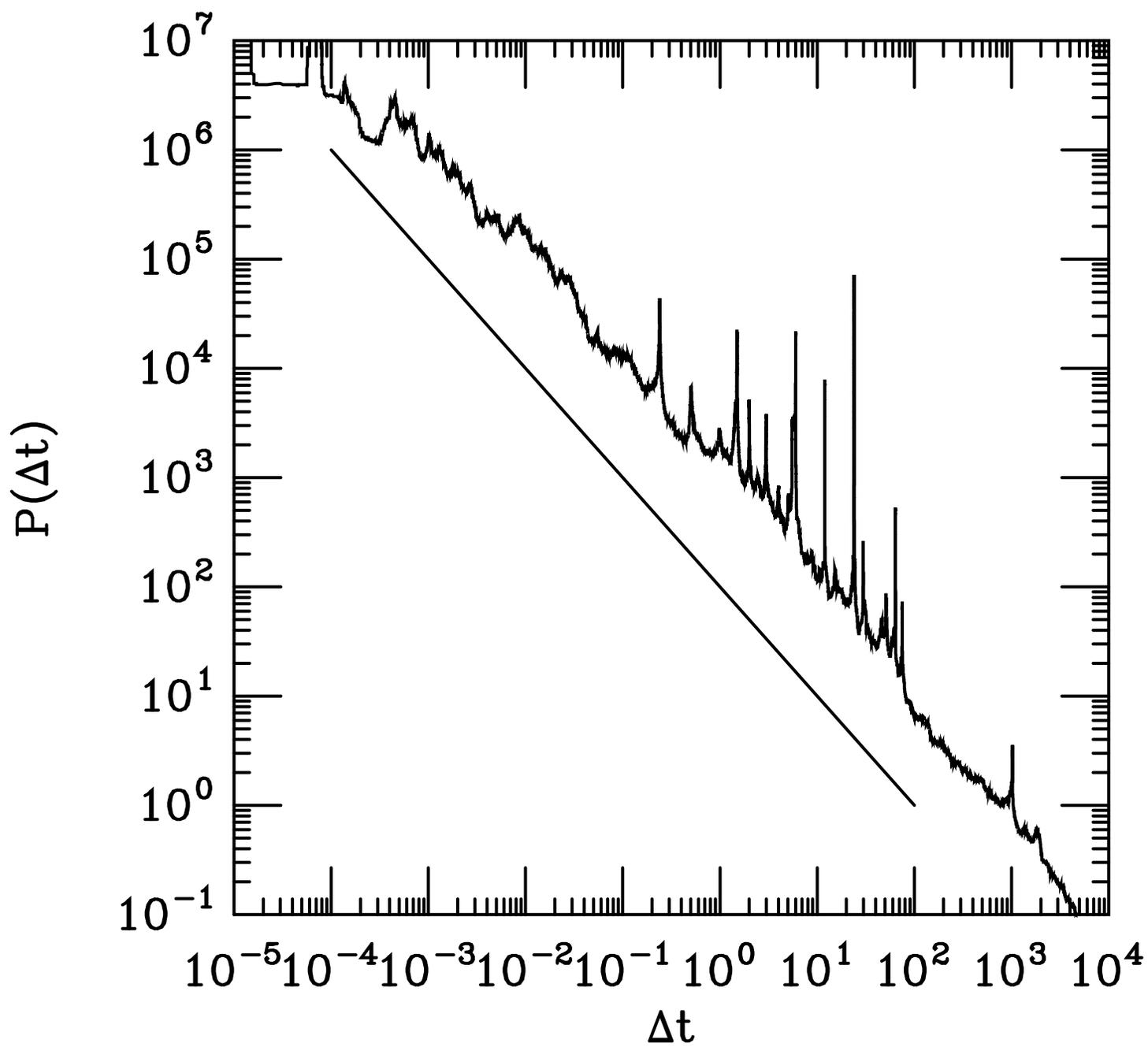
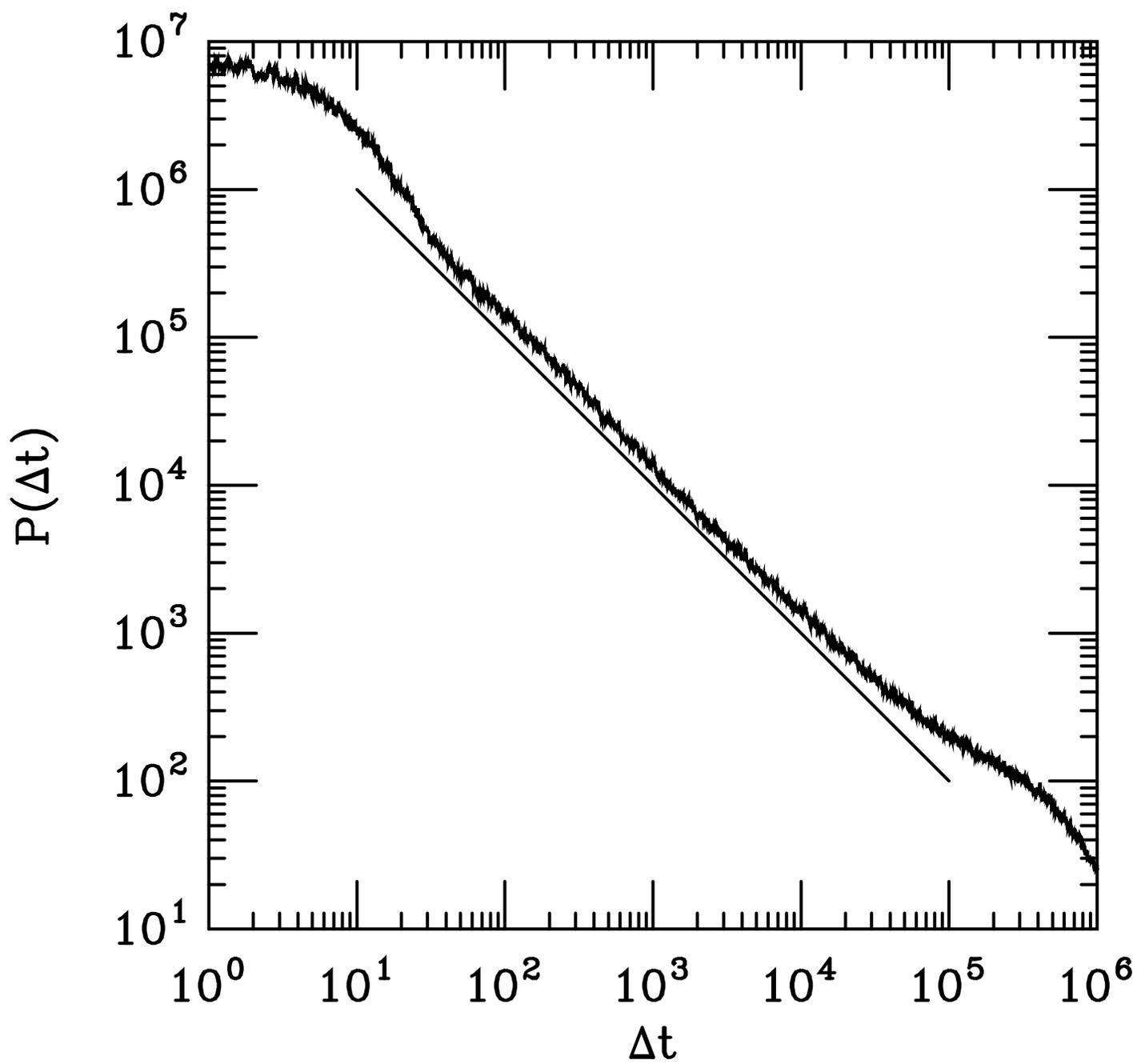
\end{document}